# Nanofluid Heat Transfer in Parallel Plates with Variable Magnetic Field

Sangeetha P S[1] and Sukanta Nayak[2]

[1]Department of Mathematics, Amrita School of Engineering, Amrita Vishwa Vidyapeetham, Coimbatore, 641112, India

[2]Department of Mathematics, School of Advanced Sciences, VIT-AP University, Amaravati, Andhra Pradesh, 522237, India

## Abstract

Uncertainties plays an important character in almost every problem which generally not considered and ideal cases are studied. As such, the consequences cost more and need model prediction. In view of these, this paper investigates nanofluid heat transfer problem under a variable magnetic field with uncertain bounded parameters and analyzed effect of uncertainness over the field variables of the system. The variations of velocity, temperature and concentration profiles with the uncertainness for the said problem are reported here. The main challenge one face is the presence of uncertainties make the system complicated to study. So, alternate idea should be developed such that one can overcome the same. Hence, the involved uncertain parameters are considered as intervals and one parametric mapping technique is used for the same. For detail analysis of the said problem a semi analytical method viz. homotopy perturbation method is adapted with the transformation technique. Here, the suction parameter, squeeze number and Hartmann numbers are considered as intervals. Then considering different combinations of these parameters the problem is solved and sensitiveness of the same for the said problem is analysed. Finally, based on the sensitiveness, a relation of between the uncertain parameters with the mentioned system has been established.

---

[1] Sangeetha P S (Email: sangeetharithu@gmail.com)
[2] Corresponding Author
Sukanta Nayak (Email: sukantgacr@gmail.com)

# 1. Introduction

Nanofluid is the mixture of base fluid and nano-scale particles. One of the major use of nanofluid in industry is to improve the heat flow rate within a system domain. In this regards, many researchers have contributed various articles to study the nanofluid flow and its effects on the system [Yu and Choi (2003), Patel et al. (2005), Minsta et al. (2009), Domairry et al. (2012), Sheikholeslami and Ganji (2014), Sheremet et al. (2014), Sheremet and Pop (2014), Garoosi et al. (2015), Garoosi and Hoseininejad (2016), Sheikholeslami and Ganji (2017)]. Hassani et al. (2011) took a stretching sheet and studied the boundary layer flow of a nanofluid. They reported that the Nusselt number is inversely proportional to Prandtl number in terms of the incearse and decrease of their values. Sheikholeslami and Ganji (2013) considered Homotopy Perturbation Method (HPM) to analyse the nanofluid flow squeezed between parallel plates. They observed that the Nusselt number, nanoparticle volume fraction, the squeeze number and Eckert number are directly related with each other when two plates are separated. Furthermore, it has an indirect relationship if squeeze number are taken into account for the case where two plates are squeezed. Hatami et. al (2014) used weighted residue method viz. Galerkin method (GM) and numerical least square method (LSM) to understand the heat flow rate of nanofluid between two parallel plates. The two plates are designed in such a way that one of the plate exerted external heat and the other one is injected with coolant fluid. Then it is noted that the Nusselt number directly proportionate to the nanoparticle volume fraction and Reynolds number. Further, Sheikholeslami et al. (2014) shown that the improvement of heat flow establish a reverse relationship with Hartmann number and Rayleigh number. Gupta and Saha Ray (2015) took Chebyshev wavelet expansion method to investigate unsteady nanofluid flow squeezing between two parallel plates. Acharya et al. (2016) the magnetic field environment to analyse the Cu-water and Cu-kerosene nanofluids squeezing flow between two parallel plates. To study the same they used numerical methods viz. Runge Kutta (RK-4) method and Differential Transformation Method (DTM).

As such, extensive research of nanofluid sqeezeing flow between two parallel plates fetch the demand and its applications made it one of the important research topic in engineering applications viz. processing of polymer, food and beverage processing, cooling towers, hydrodynamical machines, lubrication system, chemical processing and damage of crops due to freezing etc. If we see more insight of the same then we can find few more research works such as Mahmood et al. (2007) tell about the feature of nanofluid heat transfer over a porous surface, Aziz (2010) has taken a viscous fluid past an unsteady stretching sheet, Domairry and Aziz (2009) used magnetohydrodynamic (MHD) environment to analyze the squeezing flow of a viscous fluid. In this regard, Joneidi et al. (2011) also considered the magnetohydrodynamic environment for the same but both analytical and numerical approaches were adopted. Hayat et al. (2012) analyze the squeezing flow of second grade fluid for the above mentioned problem.

Besides the discussed problem, one of the most important factor viz. nonlinearty comes into play when we consider real world problems. Therefore, many numerical, semianalytical, and analytical methods are deployed. Methods such as Variational Iteration Method (VIM), Perturbation Method (PM), and Homotopy Perturbation Method (HPM) gained popularity due to its easy implementation. One of the common point in these class of methods are to eliminate a small parameter used in the same. But, one of the advantage in HPM is no need of such small parameter which made this method more compatible in computational sense. So, in this paper HPM is taken as tool to handle the problem. The HPM was introduced in He (2004) and found that a fast convergence when compared with other methods of the same class. Also, HPM has an eligibility of solving a large class of nonlinear problems with efficient. One of the advantage is that often it takes few number of iterations to achieve high accuracy solution. So, HPM is employed in many research problems of science and engineering. In this regard, many researchers have contributed various valueable works. But, here few of the important works are reported. Sheikholeslami et al. (2011) took HPM as a tool to study the rotating MHD viscous flow with stretching and porous surfaces. The study showed a inverse proportional relationship of magnetic parameter or viscosity with the Nusselt number and direct proportional relationship of rotation, blowing velocity and Prandtl number with the Nusselt number. Further, Moghimi et al. (2011) considered HPM to solve nonlinear MHD Jeffery Hamel problem and reported the field variavbles. A comprehensive analysis of viscoelastic fluid transfer with MHD in a porous medium was presented by Baag et al. (2015). Their research shown that a reverse flow behavior of high heat capacity viscoelastic fluid in the presence of magnetic field. The higher cooling of the plate in case of viscoelastic flow also causes a flow reversal. Finally, using fractional calculus researchers have started to analyse MHD viscous flow. In this respect, Asjad et. al (2020) used Caputo type fractional derivative and solved MHD viscous fluid flow through porous medium.

From the mentioned literature, it is observed that the problem is discussed only when the parameters are crisp only. Whereas, in actual practice there involve uncertainties due to coeffients, contants, parameters and boundary condition which makes the problem even more complicated and challenging to investigate [Nayak and Chakraverty (2015), Nayak and Chakraverty (2016), Nayak and Chakraverty (2018), Karunakar and Chakraverty (2018), Biswal et al. (2020), Nayak (2020), Biswal et al. (2021)]. Hence, this manuscript included the parameters as interval and then the modeled problem is investigated. The HPM is modified for interval parameters and Interval HPM (IHPM) is proposed. The same has been considered to study the nanofluid heat flow and various uncertain field parameters in presence of a variable magnetic field. Here, the parameter viz. suction parameter, and numbers viz. squeeze and Hartmann numbers are taken as intervals and the formulated problem is solved for different combinations interval numbers and parameter. Then, the obtained velocity, temperature and concentration profiles are analysed with the sensitiveness of the uncertain parameters.

2. **Nomenclature**

| | |
|---|---|
| $t$ | Time |
| $\alpha$ | Coefficient for change of magnetic field w.r.t. time |
| $\theta$ | Non-dimension temperature profile |
| $\phi$ | Non-dimension nanoparticles concentration |
| $\eta$ | Non-dimension height |
| $A$ | Suction/blowing parameter |
| $B$ | Magnetic field |
| $C_w$ | Nanoparticles concentration |
| $H$ | Height between the plate |
| $k$ | Thermal conductivity |
| $Le$ | Lewis number |
| $M$ | Hartmann number |
| $Nb$ | Brownian motion coefficient |
| $Nt$ | Thermophoretic coefficient |
| $Nu$ | Nusselt number |
| $p$ | Pressure |
| $pr$ | Prandtl number |
| $q$ | Transferred heat coefficient |
| $r$ | Radius direction |
| $S$ | Squeeze number |

| $T_w$ | Temperatures of Nanoparticles |
|---|---|
| $u, w$ | Velocities in $r$ and $z$ directions |
| $z$ | Vertical direction |
| $\sigma$ | Stephan-Boltzmann coefficient |
| $\rho_f$ | Fluid density |

### 3. Interval Homotopy Perturbation Method (IHPM)

The idea of IHPM is the hybridization of the concept of interval analysis with the Homotopy Perturbation Method (HPM). The purpose of interval analysis is to handle uncertain parameters involved in the system which are considered as interval. But, the traditional interval arithmetic is very much challenging to compute the intervals. So, here we have considered a transformation technique which converts the intervals to single variable functions and then the same is used for computation. For the sake of completeness first the traditional interval arithmetic is discussed. Then the transformation technique is introduced. Finally, HPM is explained precisely in this section.

An interval $A$ can be defined as

$$A = [\alpha_1, \alpha_2] = \{t | \alpha_1 \leq t \leq \alpha_2; \alpha_1, \alpha_2 \in \mathcal{R}\}, \quad (1)$$

where $\alpha_1 \leq a_2$. If $\alpha_1 = \alpha_2$, then $A$ is said to be crisp.

Two intervals are equal only when the the left endpoint and right endpoint of the intervals are the same. For illustration we may take two arbitrary intervals $A = [\alpha_1, \alpha_2]$ and $B = [\beta_1, \beta_2]$, then they are said to be equal if $\alpha_1 = \beta_1$ and $\alpha_2 = \beta_2$.

The width of an interval $A$ is defined by $w = \alpha_2 - \alpha_1$. Whereas, the midpoint of $A$ is $m = \frac{\alpha_1 + \alpha_2}{2}$.

To compute interval uncertainties, the traditional interval arithmetic operations are defined as

$$[\alpha, \alpha_2] + [\beta_1, \beta_2] = [\alpha_1 + \beta_1, \alpha_2 + \beta_2]; \quad (2)$$

$$[\alpha_1, \alpha_2] - [\beta_1, \beta_2] = [\alpha_1 - \beta_2, \alpha_2 - \beta_1]; \quad (3)$$

$$[\alpha_1, \alpha_2] \times [\beta_1, \beta_2] = [\min(\alpha_1\beta_1, \alpha_1\beta_2, \alpha_2\beta_1, \alpha_2\beta_2), \max(\alpha_1\beta_1, \alpha_1\beta_2, \alpha_2\beta_1, \alpha_2\beta_2)]; \quad (4)$$

$$\frac{[\alpha_1, \alpha_2]}{[\beta_1, \beta_2]} = \left[\min\left(\frac{\alpha_1}{\beta_1}, \frac{\alpha_1}{\beta_2}, \frac{\alpha_2}{\beta_1}, \frac{\alpha_2}{\beta_2}\right), \max\left(\frac{\alpha_1}{\beta_1}, \frac{\alpha_1}{\beta_2}, \frac{\alpha_2}{\beta_1}, \frac{\alpha_2}{\beta_2}\right)\right], \text{ where } \beta_1, \beta_2 \neq 0. \quad (5)$$

Alternatively, an interval $A = [\alpha_1, \alpha_2]$ can be written as the following crisp form by using a parameter $\alpha, 0 \leq \alpha \leq 1$.

$$A = f(\alpha) = \alpha(\alpha_2 - \alpha_1) + \alpha_1 \quad (6)$$

Substituting, $\alpha = 0$ and $\alpha = 1$, one may get the left and right bounds of the interval $A$ respectively. The expression Eq. (6) converts the interval into a sigle variable function which is used here to compute the interval uncertainties.

The basic idea of HPM can be understood from the below mentioned differential equation.

Let us consider the differential equation

$$A(T) - f(r) = 0, r \in \Omega \tag{7}$$

with the boundary condition

$$B\left(T, \frac{dV}{d\eta}\right) = 0, r \in \Gamma \tag{8}$$

where $A$ is a differential operator, $B$ is the boundary operator, $f(r)$ is a known analytical function and $\Gamma$ is the boundary of the domain $\Omega$.

The operator $A$ can be classified into two parts viz. linear ($L$) and non-linear ($N$). Then the Eq. (7) can be written as

$$L(T) + N(T) - f(r) = 0, \quad r \in \Omega \tag{9}$$

One may construct a homotopy $v(r, q): \Omega \times [0,1] \to R$ satisfying

$$H(v, q) = (1 - q)[L(v) - L(T_0)] + q[A(v) - f(r)] = 0 \tag{10}$$

where $q$ is an embedding parameter lies in between 0 and 1, $T_0$ is an initial approximation satisfying boundary condition Eq. (9).

From Eq. (10), it is observed that when $q = 0; L(v) = L(T_0)$ and for $q = 1; A(v) - f(r) = 0$. In other words, when $q$ converges to 1, we get the solution of Eq. (7).

Since, $q$ is a small parameter, the solution of Eq. (10) can be expressed as in terms of power series in $q$

$$v = v_0 + qv_1 + q^2 v_2 + q^3 v_3 + \cdots \tag{11}$$

By setting $q = 1$, one may get the following solution for Eq. (7)

$$v = \lim_{q \to 1}(v_0 + qv_1 + q^2 v_2 + q^3 v_3 + \cdots) \tag{12}$$

## 4. Problem description

Consider the nanofluid flow problem shown in Fig. 1. Here, the incompressible sqeezing nanofluid is taken in parallel disks with the a distance between the parallel disks is $h(t) = H(1 - at)^{-1/2}$. The sqeeze flow is studied under variable magnetic field environment having a magnetic field strength $B(t) = B_0(1 - at)^{-1/2}$. The magnetic field is suuplied perpendicular to the parallel disks. Sketch of the fluid flow directions inside the movable parallel plates are shown in Fig. 1 for easy illustration of the problem. The left half fluid flow of the channel represent axisymmetric squeezing flow. Here, '*' marks are indicating the magnetic field which is normal to the plane, where positive and negative charges are denoted by '+' and '-' symbols. The positive and negative charges, Hall effect and the particles are subjected to Lorentz force. The Lorentz force and flow direction makes an obtuse angle with each other.

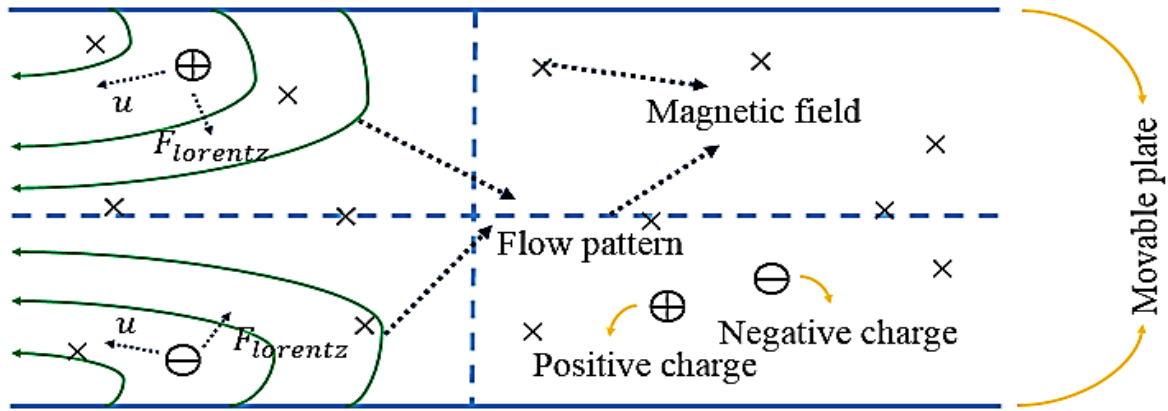

Fig.1 Schematic diagram of nanofluid flow within a parallel disks under variable magnetic field environment

Nanofluid flow experiences difference in temperatures and concentration levels at lower and upper disks. The temperatures at lower and upper disks are denoted as $T_w$ and $T_h$ respectively. Whereas, the nanoparticles concentration at lower and upper disks are denoted as $C_w$ and $C_h$ respectively. The lower disk posseses stationary state and the upper disk marches towards it at $z = h(t)$ with a velocity of $\frac{dh}{dt}$. Depending on the value $a$ either the disks (plates) are squeezes or separated. For $a < 0$, the two disks (plates) are separated and $a > 0$, the two disks are squeezed and touches each other at $t = \frac{1}{a}$. Here, the viscous dissipation effect, the generation of heat due to friction caused by shear in the flow, is preserved. This provide more impact if large viscous fluid flows or speed of the same is high. It is pre-assumed that the nanofluid is a mixture of two components with the properties viz. incompressible, no-chemical reaction, negligible viscous dissipation, negligible radiative heat transfer, nano-solid-particles and the base fluid are in thermal equilibrium and no slip occurs between them. Then, the governing equations of the flow and mass transfer in viscous fluid are

$$\rho \left(\frac{\partial \vec{V}}{\partial t} + (\vec{V}.\vec{\nabla})\vec{V}\right) = -\vec{\nabla}P + \mu \nabla^2 \vec{V} + \sigma(\vec{J} \times \vec{B}), \tag{13}$$

$$(\rho C_p)\left(\frac{\partial T}{\partial t} + (\vec{V}.\vec{\nabla})T\right) = k\nabla^2 T + \tau[D_s(\vec{\nabla}T.\vec{\nabla}C) + \frac{D_T}{T_m}(\vec{\nabla}T.\vec{\nabla}T)], \tag{14}$$

$$(\rho C_p)\left(\frac{\partial C}{\partial t} + \vec{V}.\vec{\nabla}\right)C\right) = D_B \nabla^2 C + \frac{D_T}{T_m}\nabla^2 T, \tag{15}$$

$$\vec{\nabla}.\vec{V} = 0, \tag{16}$$

where $\vec{V} = (u, v, w)$ is the velocity vector and $\vec{J} = \vec{E} + (\vec{V} \times \vec{B})$. The small magnetic field nullify the Reynolds number, $\vec{E}$, so $\vec{J} = (\vec{V} \times \vec{B})$.

The operator $\vec{\nabla}$ is defined as

$$\vec{\nabla} = \left(\frac{\partial}{\partial x}, \frac{\partial}{\partial y}, \frac{\partial}{\partial z}\right). \tag{17}$$

The boundary conditions are

$$\text{At } z = 0, \text{ we have, } u = 0 \quad w = \frac{w_0}{\sqrt{1-at}}, \quad T = T_w, \quad C = C_w$$
$$\text{At } z = h(t), \text{ we have, } u = 0, \quad w = \frac{dh}{dt}, \quad T = T_h, \quad C = C_h \tag{18}$$

The last term in the engery equation is the total diffusion mass flux for nanoparticles which give the sum of two diffusion terms (Brownian motion and thermophoresis), where $\tau$ is the dimensionless parameter between the ratio of effective heat capacity of the nanoparticle materials to heat capacity of the fluid. Thus, the value of $\tau$ is different for different fluids and nanoparticle materials. Following are the similar transformation equations which will be used in Eq. (13) and (14) to eliminate pressure gradient.

$$u = \frac{ar}{2(1-at)} f'(\eta), w = -\frac{aH}{\sqrt{1-at}} f(\eta), \eta = \frac{z}{H\sqrt{1-at}}$$
$$B(t) = \frac{B_0}{\sqrt{1-at}}, \theta = \frac{T - T_h}{T_w - T_h}, \emptyset = \frac{C - C_h}{C_w - C_h} \tag{19}$$

After eliminating pressure gradient from the resulting equation, the Eq. (15) can be rewritten and the following coupled system of nonlinear equations [Hatami et. al (2015)] are obtained.

$$f'''' - S(\eta f''' + 3f'' - 2ff''') - M^2 f'' = 0$$
$$\theta'' + PrS(2f\theta' - \eta\theta') + PrNb\theta'\emptyset' + PrNt\theta'^2 = 0 \tag{20}$$
$$\emptyset'' + LeS(2f\theta' - \eta\emptyset') + \frac{Nt}{Nb}\theta'' = 0$$

with the boundary conditions

$$f(0) = A, \quad f'(0) = 0, \quad \theta(0) = \emptyset(0) = 1$$
$$f(1) = \frac{1}{2}, \quad f'(1) = 0, \quad \theta(1) = \emptyset(1) = 0 \tag{21}$$

Thus, the continuity equation is satisfied and where $A > 0$ denotes the suction of fluid in the lower disk and $A < 0$ indicates the injection flow. In terms of conditions given in Eq.(18), one may write

$$Nu^* = (1-at)^{1/2}, Nu = -\theta'(1) \tag{21}$$

Next section is dedicated for application of developed IHPM to solve the above problem.

## 5. Implementation of IHPM

Construct a homotopy with interval parameters and suppose that the solution of Eq. (7) has the form

$$\widetilde{H}(\tilde{f}, q) = (1 - q)\left(\tilde{f}^{iv} - \tilde{f}_0^{iv}\right) + q(\tilde{f}''' - S(\eta\tilde{f}''' + 3\tilde{f}'' - 2\tilde{f}\tilde{f}''')M^2\tilde{f}'') = 0 \quad (22)$$

$$\widetilde{H}(\tilde{\theta}, q) = (1 - q)(\tilde{\theta}'' - \tilde{\theta}_0'') + q(\tilde{\theta}'' + PrS(2\tilde{f}\tilde{\theta}' - \eta\tilde{\theta}') + PrNb\tilde{\theta}'\widetilde{\emptyset}' + PrNt\tilde{\theta}'^2) = 0 \quad (23)$$

$$\widetilde{H}(\widetilde{\emptyset}, q) = (1 - q)\left(\widetilde{\emptyset}'' - \widetilde{\emptyset}_0''\right) + q\left(\widetilde{\emptyset}'' + LeS(2\tilde{f}\widetilde{\emptyset}' - \eta\widetilde{\emptyset}') + \frac{Nt}{Nb}\tilde{\theta}''\right) = 0 \quad (24)$$

We consider $\tilde{f}, \tilde{\theta}$ and $\widetilde{\emptyset}$ as follows

$$\tilde{f}(\eta) = \tilde{f}_0(\eta) + \tilde{f}_1(\eta) + \tilde{f}_2(\eta) + \cdots = \sum_{i=0}^{n} \tilde{f}_i(\eta)$$

$$\tilde{\theta}(\eta) = \tilde{\theta}_0(\eta) + \tilde{\theta}_1(\eta) + \tilde{\theta}_2(\eta) + \cdots = \sum_{i=0}^{n} \tilde{\theta}_i(\eta) \quad (25)$$

$$\widetilde{\emptyset}(\eta) = \widetilde{\emptyset}_0(\eta) + \widetilde{\emptyset}_1(\eta) + \widetilde{\emptyset}_2(\eta) + \cdots = \sum_{i=0}^{n} \widetilde{\emptyset}_i(\eta)$$

Substituting the $\tilde{f}, \tilde{\theta}$ and $\widetilde{\emptyset}$ from Eqs. (25) to (27) into Eqs. (8) to (10) and collecting the various power of $q$-terms, we have

For $q^0$:

$$\tilde{f}_0^{iv} = 0$$
$$\tilde{\theta}_0'' = 0 \quad (26)$$
$$\widetilde{\emptyset}_0'' = 0$$

and the boundary conditions

$$\tilde{f}_0(0) = A, \quad \tilde{f}_0'(0) = 0, \quad \tilde{\theta}_0(0) = \widetilde{\emptyset}_0(0) = 1$$
$$\tilde{f}_0(1) = 1/2, \quad \tilde{f}_0'(1) = 0, \quad \tilde{\theta}_0(1) = \widetilde{\emptyset}_0(1) = 0 \quad (27)$$

For $q^1$:

$$\tilde{f}_1^{iv} - S\eta\tilde{f}_0''' - 3S\tilde{f}_0'' + 2S\tilde{f}_0\tilde{f}_0'' - M^2\tilde{f}_0'' = 0$$
$$\tilde{\theta}_1'' + 2PrS\tilde{f}_0\tilde{\theta}_0' - PrS\eta\tilde{\theta}_0' + PrNb\tilde{\theta}_0'\widetilde{\emptyset}_0' + PrNt\tilde{\theta}_0'^2 = 0 \quad (28)$$
$$\widetilde{\emptyset}_1'' + 2LeS\tilde{f}_0\widetilde{\emptyset}_0' - LeS\eta\widetilde{\emptyset}_0' + \frac{Nt}{Nb}\tilde{\theta}_0'' = 0$$

and the boundary conditions

$$f_1(0) = 0, \quad f_1'(0) = 0, \quad \theta_1(0) = \phi_1(0) = 0$$
$$f_1(1) = 0, \quad f_1'(1) = 0, \quad \theta_1(1) = \phi_1(1) = 0 \tag{29}$$

By solving equations (25) and (27) with boundary conditions, we get

$$\tilde{f}_0(\eta) = 0.166667(-6 + 12A)\eta^3 + 0.5(3 - 6A)\eta^2 + A$$
$$\tilde{\theta}_0(\eta) = -\eta + 1 \tag{30}$$
$$\tilde{\phi}_0(\eta) = -\eta + 1$$

Similarly,

$$\begin{aligned}\tilde{f}_1(\eta) = {}& .4000008AS\eta^5 - .2000004S\eta^5 + .375S\eta^4 - .01428S\eta^7 \\ & - .0571SA^2\eta^7 + .0571SA\eta^7 - .25AS\eta^4 + .0499S\eta^6 - .2SA\eta^6 \\ & + .2SA^2\eta^6 - .999A^2S\eta^4 - .05M^2\eta^5 + .099AM^2\eta^5 + .12M^2\eta^4 \\ & - .1857SA\eta^3 - .2785S\eta^3 - .25AM^2\eta^4 - .849937M^2\eta^3 \\ & + 1.485SA^2\eta^3 + .1785SA\eta^2 + .2AM^2\eta^3 - .05M^2A\eta^2 \\ & - .0628SA^2\eta^2 + .0678S\eta^2 + .025M^2\eta^3 \end{aligned} \tag{31}$$

$$\begin{aligned}\tilde{\theta}_1(\eta) = {}& -.1PrS\eta^5 + .2PrSA\eta^5 + .25PrS\eta^4 - .5PrSA\eta^4 - .5PrNb\eta^2 \\ & - .5PrNt\eta^2 - .1666667PrS\eta^3 + .0166672PrS\eta + .5PrNb\eta \\ & + .5PrNt\eta - .7PrSA\eta + PrSA\eta^2\end{aligned}$$

$$\tilde{\phi}_1(\eta) = LeS[.1000002\eta^5 + .2A\eta^5 + .25\eta^4 - .5A\eta^4 + A\eta^2 - .166666\eta^3 + .0166662\eta - .7A\eta]$$

Substitute the above solutions in Eq. (25), we get

$$\begin{aligned}\tilde{f}(\eta) = {}& 0.166667(-6 + 12A)\eta^3 + 0.5(3 - 6A)\eta^2 + A \\ & + [.4000008AS\eta^5 - .2000004S\eta^5 + .375S\eta^4 - .01428S\eta^7 \\ & - .0571SA^2\eta^7 + .0571SA\eta^7 - .25AS\eta^4 + .0499S\eta^6 - .2SA\eta^6 \\ & + .2SA^2\eta^6 - .999A^2S\eta^4 - .05M^2\eta^5 + .099AM^2\eta^5 + .12M^2\eta^4 \\ & - .1857SA\eta^3 - .2785S\eta^3 - .25AM^2\eta^4 - .849937M^2\eta^3 \\ & + 1.485SA^2\eta^3 + .1785SA\eta^2 + .2AM^2\eta^3 - .05M^2A\eta^2 \\ & - .0628SA^2\eta^2 + .0678S\eta^2 + .025M^2\eta^3] + \cdots \end{aligned} \tag{32}$$

$$\begin{aligned}\tilde{\theta}(\eta) = {}& [-\eta + 1] \\ & + [-.1PrS\eta^5 + .2PrSA\eta^5 + .25PrS\eta^4 - .5PrSA\eta^4 - .5PrNb\eta^2 \\ & - .5PrNt\eta^2 - .1666667PrS\eta^3 + .0166672PrS\eta + .5PrNb\eta \\ & + .5PrNt\eta - .7PrSA\eta + PrSA\eta^2] + \cdots\end{aligned}$$

$$\tilde{\phi}(\eta) = [-\eta + 1] + [LeS[.1000002\eta^5 + .2A\eta^5 + .25\eta^4 - 0.5A\eta^4 + A\eta^2 - .166666\eta^3 + .0166662\eta - .7A\eta]] + \cdots$$

## 6. Results and discussion

The above discussed nanofluid heat transfer problem in presence of variable magnetic fluid with uncertain parameters viz. squeeze number ($S$), suction parameter ($A$) and Hartmann number ($M$) are considered here. The uncertain $S$, $A$, and $M$ with $\pm 5\%$ of error of uncertainness are taken into account for investigation. The investigation is divided into the following three steps.

Step 1

In this step the velocity profile of the nanofluid in presence of magnetic field is studied. Three combinations of pair of interval uncertainties are taken and the velocity is observed. The three combinations are (i) $S$ and $M$ are intervals with constant $A$ value, (ii) $S$ and $A$ are intervals with constant $M$ value, and (iii) $A$ and $M$ are intervals with constant $S$ value. The numerical values of velocity profiles for (i), (ii), and (iii) are shown in Figs. 2 to 4.

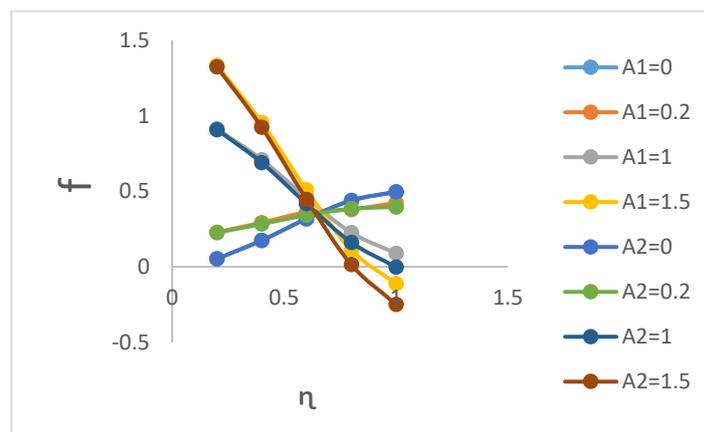

Fig. 2. Velocity profile when $S = [0.95, 1.05]$ and $M = [0.95, 1.05]$

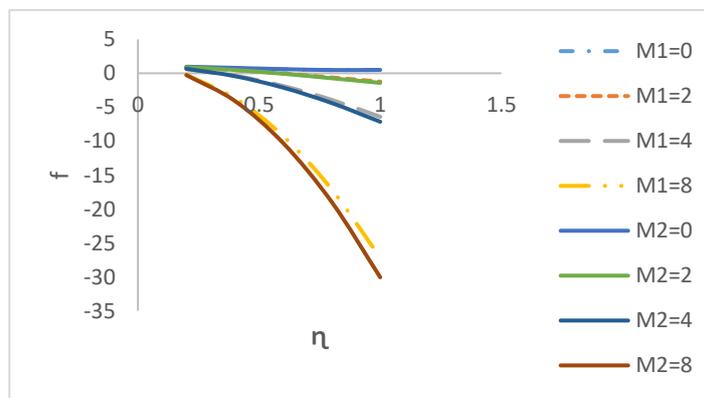

Fig. 3. Velocity profile when $A = [0.95, 1.05]$ and $S = [0.95, 1.05]$

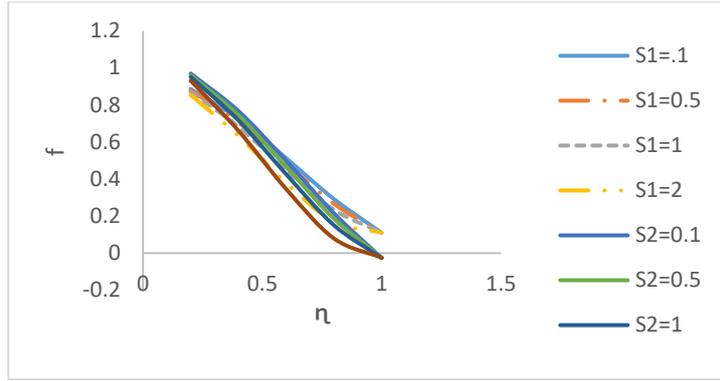

Fig. 4. Velocity profile when $A = [0.95, 1.05]$ and $M = [0.95, 1.05]$

Based on the comparison of uncertain width in the above three cases, it has been seen that when $A$ and $M$ are considered as intervals with constant $S$ value the uncertain width of the velocity is more compared to the other cases. Hence, it is noted that the velocity profile is more sensitive for the values of suction parameter $A$ and Hartmann number $M$. This suggests that a slight variation in terms of uncertainness for suction parameter and Hartmann number may increases the uncertainness of velocity of the nanofluid when it is considered in presence of variable magnetic field. As such, the same can be taken care with priority when it will be used in real practice.

Step 2

In this step the temperature profile of the nanofluid in presence of magnetic field is studied. Three combinations of pair of interval uncertainties are taken and the temperature is observed. The three combinations are (i) $S$ and $M$ are intervals with constant $A$ value, (ii) $S$ and $A$ are intervals with constant $M$ value, and (iii) $A$ and $M$ are intervals with constant $S$ value. The numerical values of temperature profiles for (i), (ii), and (iii) are shown in Figs. 5 to 7.

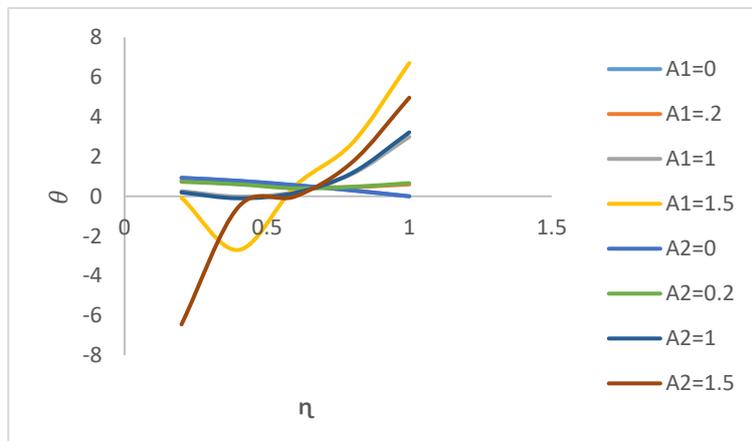

Fig. 5. Temperature profile when $S = [0.95, 1.05]$ and $M = [0.95, 1.05]$

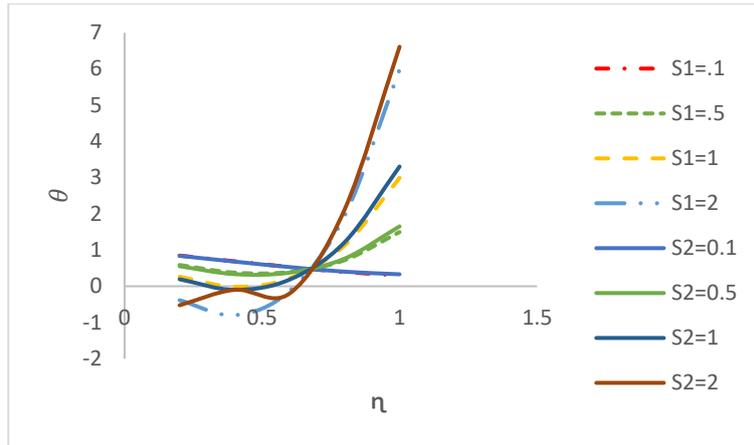

Fig. 6. Temperature profile when $A = [0.95, 1.05]$ and $M = [0.95, 1.05]$

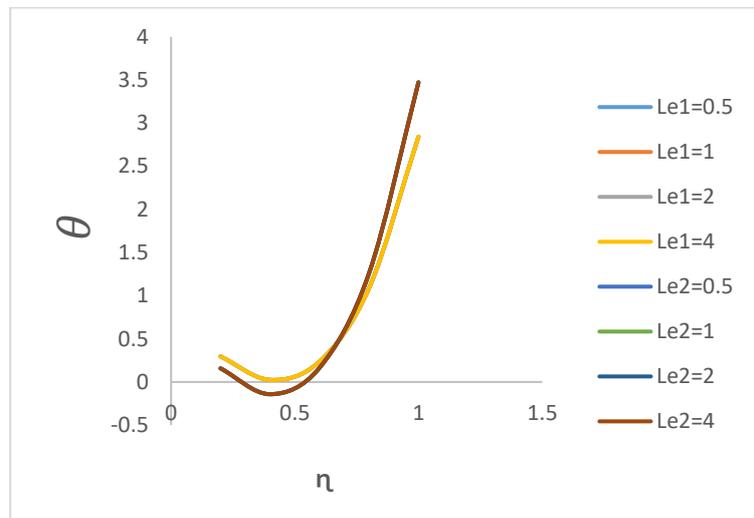

Fig. 7. Temperature profile when $A = [0.95, 1.05]$ and $S = [0.95, 1.05]$

Based on the comparison of uncertain width in the above three cases, it has been seen that when $A$ and $S$ are considered as intervals with constant $M$ value the uncertain width of the temperature is more compared to the other cases. Hence, it is noted that the temperature profile is more sensitive for the values of suction parameter $A$ and sqeeze number $S$. This suggests that a slight variation in terms of uncertainness for suction parameter and squeeze number may increases the uncertainness of temperature of the nanofluid when it is considered in presence of variable magnetic field. As such, the same can be taken care with priority when it will be used in real practice.

Step 3

In this step the concentration profile of the nanofluid in presence of magnetic field is studied. Three combinations of pair of interval uncertainties are taken and the concentration is observed. The three combinations are (i) $S$ and $M$ are intervals with constant $A$ value, (ii) $S$ and $A$ are intervals with constant $M$ value, and (iii) $A$ and $M$ are intervals with constant $S$ value. The numerical values of concentration profiles for (i), (ii), and (iii) are shown in Figs. 8 to 10.

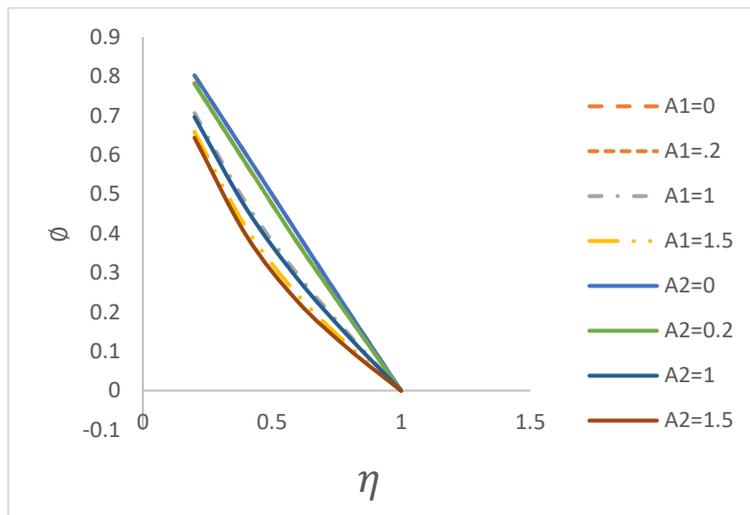

Fig. 8. Concentration profile when $S = [0.95, 1.05]$ and $M = [0.95, 1.05]$

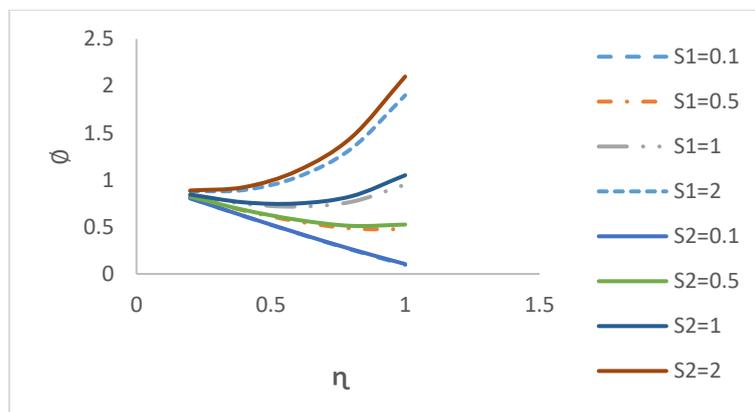

Fig. 9. Concentration profile when $A = [0.95, 1.05]$ and $M = [0.95, 1.05]$

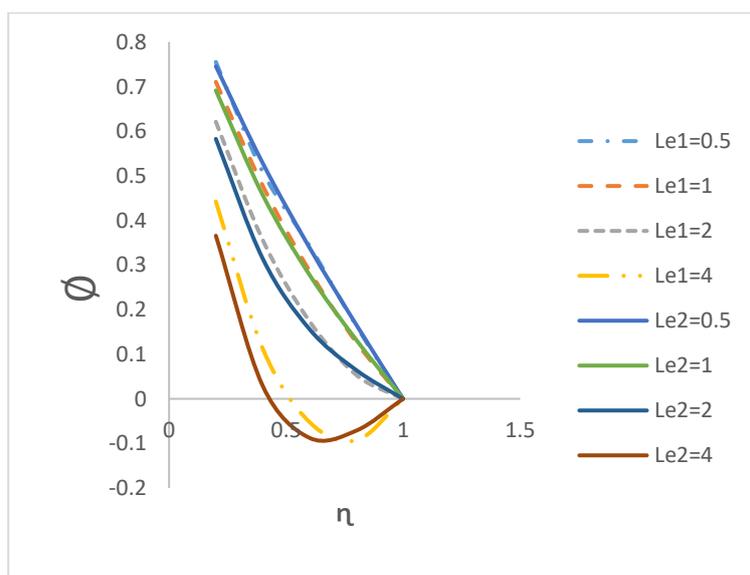

Fig. 10. Concentration profile when $A = [0.95, 1.05]$ and $S = [0.95, 1.05]$

Based on the comparison of uncertain width in the above three cases, it has been seen that when $A$ and $S$ are considered as intervals with constant $M$ value the uncertain width of the concentration is more compared to the other cases. Hence, it is noted that the concentration profile is more sensitive for the values of suction parameter $A$ and sqeeze number $S$. This suggests that a slight variation in terms of uncertainness for suction parameter and squeeze number may increases the uncertainness of concentration of the nanofluid when it is considered in presence of variable magnetic field. As such, the same can be taken care with priority when it will be used in real practice.

## 7. Conclusion

Nanofluid plays a vital role in heat transfer problems due to its variable impact of change in temperature. Taking this in account in this paper additional magnatic field environment has been considered with the nanofluid heat transfer. To understand the behavior of heat transfer and realistic study of the same, suction parameter, sqeeze number and Hartmann numbers are taken as uncertain with 5 percent error. Then the problem is investigated by using a semianalytical method viz. interval homotopy perturbation method. The variation of velocity, concentration, and temperature profile is reported with uncertain environment. Further, it is noted that if the suction parameter and Hartman number posses little margin of error with constant squeeze number then the velocity changes drastically. If suction parameter and sqeeze number posses little margin of error with constant Hartmann number then both the temperature and concentration changes drastically. Finally, the model and observed results may be refered in the industrial reference for enhancement of temperature, velocity and concentration profile.

**Data availability statement**

This manuscript has no associated data.

**Conflict of interest statement**

The Authors' have no conflict of interest. This is the authors' original work and has not been published nor it has been submitted simultaneously elsewhere.